\documentclass{ws-ijmpa}
\usepackage[super,compress]{cite}
\usepackage{graphicx}
\def\draftnote{}
\begin{document}
\markboth{Y. Shi}{Beauty and Physics }
%
\catchline{29}{17}{2014}{1475001}{10} 
%

\title{Beauty and Physics:\\ 13 Important Contributions of Chen Ning Yang\footnote{A different version of this article, in Chinese, has appeared in {\em Physics}, {\bf 43} (1), 57 (2014).}
}

\author{Yu Shi}

\address{Department of Physics, Fudan University, Shanghai 200433, China\\ yushi@fudan.edu.cn }

\maketitle

\begin{history}
\received{21 April 2014}
\accepted{21 April 2014}
\published{13 June 2014}
\end{history}

\begin{abstract}
In 2012,  Chen Ning Yang received a 90th birthday gift  in the form of a black cube inscribed with his 13 most important contributions, which cover four major areas of physics:
statistical mechanics, condensed matter physics, particle physics and field theory. We briefly describe these 13 contributions and make general comments about Yang's distinctive style as a trailblazing leader in research.

\keywords{Chen Ning Yang; Yang-Mills theory; gauge theory; symmetry; particle physics; statistical mechanics; off-diagonal long-range order; history of physics.}
\end{abstract}

\ccode{PACS numbers: 01.65.+g, 11.15.-q, 05.30.-d}


\section{Introduction}

In 1928, Chen Ning Yang, aged 6, searched for sea shells on the beach. He always set his eyesight on beautiful and elegant ones. His unique style and taste since young has served to make him ``the preeminent stylist''~\cite{dyson1}, a ``leading bird''~\cite{dyson2}  and an architect of  theoretical physics in the second half of 20th century~\cite{sp,sp2,zhang,shuguangji,list,goldhaber,yang70,yang85}.

\begin{figure}
\includegraphics*{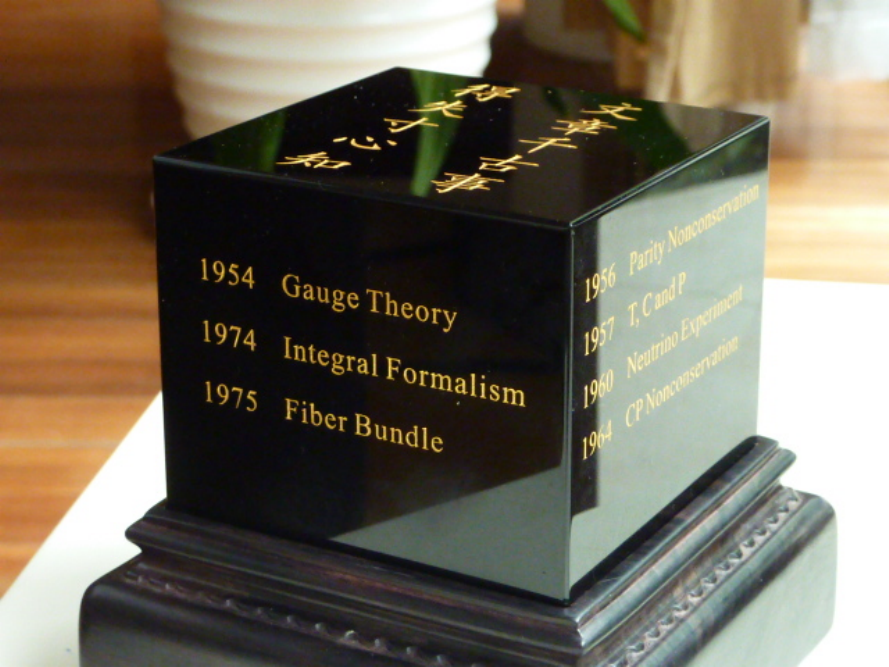} \includegraphics*{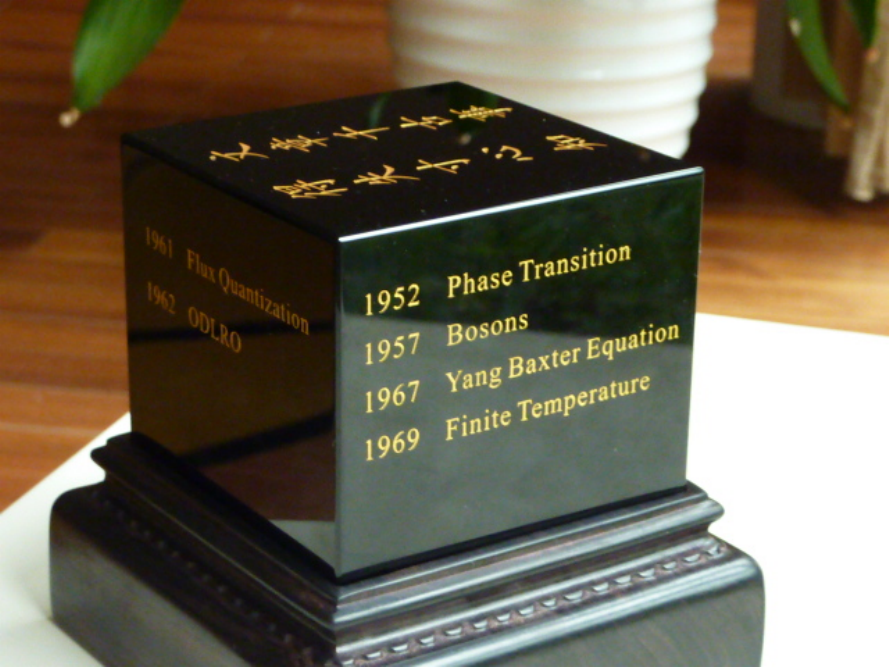}
\caption{   The black cube engraved with C. N. Yang's 13 contributions in four areas of physics. Top: front and right sides. Bottom:  back and left sides.  } \label{fig1}
\end{figure}

In 2012, Chen Ning Yang, aged 90, received a special birthday present, a black  cube of 8 cm $\times$ 8 cm $\times$ 6.6 cm (Fig.~\ref{fig1}). On its bottom is engraved ``Congratulations/on Professor Chen Ning Yang's 90th birthday/Tsinghua University'' in Chinese. On the  top is engraved two lines from  Tu Fu (712-770) in Chinese:  ``A piece of literature/Is meant for the millennium/But its ups and downs are known/Already in the author's heart.''  On the four vertical surfaces are engraved,   respectively in  clockwise order,  his 13 eminent contributions in four areas of physics:  statistical mechanics, condensed matter physics, particle physics and field theory.   One is reminded of the ``Ten Commandments'' of L. D. Landau~\cite{landau}.

The two lines from Tu Fu, which were written at Tu's old ages as a preface to the collection of his poems,  were among Yang's favorites. Yang quoted them in the preface to his {\em Selected Papers} of 1983 \cite{sp}, and he alluded to them in his  Chinese poem {\em On the Chern Class}~\cite{chern}.

Following is the list of the 13 seminal contributions in four areas, with references to the original papers.

\begin{itemize}
\item[A.]  Statistical mechanics:
\begin{itemize}
\item[A.1.] 1952 Phase Transition~\cite{52a,52b,52c}
\item[A.2.] 1957 Bosons~\cite{57h,57i,57q}
\item[A.3.] 1967 Yang-Baxter Equation~\cite{67e}
\item[A.4.] 1969 Finite Temperature~\cite{69a}
\end{itemize}
\item[B.] Condensed matter physics:
\begin{itemize}
\item[B.1.] 1961 Flux Quantization~\cite{61c}
\item[B.2.] 1962 ODLRO~\cite{62j}
\end{itemize}
\item[C.] Particle physics:
\begin{itemize}
\item[C.1.] 1956 Parity Nonconservation~\cite{56h}
\item[C.2.] 1957 T, C and P~\cite{57e}
\item[C.3.] 1960 Neutrino Experiment~\cite{60d}
\item[C.4.]  1964 CP Nonconservation~\cite{64f}	
\end{itemize}
\item[D.] Field theory:
\begin{itemize}
\item[D.1.] 1954 Gauge Theory~\cite{54b,54c}
\item[D.2.] 1974 Integral Formalism~\cite{74c}
\item[D.3.] 1975 Fiber Bundle~\cite{75c}
\end{itemize}
\end{itemize}

\section{Commentaries}

\subsection*{A.1. Phase transitions}

Statistical mechanics is one of the major research fields of Yang, in which his work is distinguished by originality, elegance, power and physical relevance.    In 1952, Yang published a paper on the spontaneous magnetization of the two-dimensional Ising model~\cite{52a}, an absolute {\it tour de force}, which F. Dyson had described as ``a  virtuoso exercise in the theory of Jacobian elliptic functions''~\cite{exp1}. The paper and its generalization by C. H. Chang gave the first indication of what is now called the universality of critical exponents~\cite{exp2}. In two papers Yang and T. D. Lee extended the study to a lattice gas model, for which the Maxwell diagram of liquid-gas transition was rigorously computed~\cite{52b,52c}. In this work,  they discovered a unit circle theorem,  which  Yang later described as a ``minor gem''~\cite{exp3} and D. Ruelle  recently used as an example in explaining how mathematical theorems are conjectured and proved~\cite{ruelle}. Furthermore, Yang and Lee showed rigorously the subtlety of the limiting process in the the theory of liquid-gas transitions, thereby settling once and for all the confusing  debate, originating from the J. E. Mayor's 1937 paper~\cite{mayor}, about ``how can the gas molecules `know' when they have to coagulate to form a liquid or solid?''~\cite{bf}

\subsection*{A.2. Bosons}

In the mid-1950s,  with his collaborators Kerson Huang, T. D. Lee and J. M. Luttinger, Yang worked on the superfluidity of liquid helium. They chose to study a mathematically well defined model, the dilute hard sphere Bosons. Through a subtle use of E.~Fermi's pseudopotential idea, they arrived at the first two terms of an asymptotic expansion of the sound velocity  of the Bosons at low temperatures~\cite{57h,57i,57q}. At the time, this work was noted for its physical and mathematical sophistication, but had absolutely no experimental relevance. Amazingly, 50 years later, with marvelous new cooling technology, this asymptotic expansion was experimentally confirmed!~\cite{altmeyer,navon}

In 1992, when asked about ``his ability to choose problems which became important ten or twenty years later'', Yang answered: ``one must look for topics that have {\em direct connections} with physical phenomenon or with the basic structure of physics.''~\cite{li}  His early choice of the Boson problem is a good example of this prescript.

\subsection*{A.3. Yang-Baxter equation}

In 1960s, his interest in off-diagonal long-range order (ODLRO) led Yang to search for models with such long-range orders. This search resurrected his interest in Bethe's Hypothesis, and in turn led him to this work~\cite{67e}, which opened the door to two  important developments: (1) the  Yang-Baxter equation,  (2) the theory of one-dimensional Fermions.  The former has now developed into a major area of research in mathematics and theoretical physics, while the latter provides now the basis for analyzing many one-dimensional  cold atom experiments~\cite{moritz}.

\subsection*{A.4. Exact solution of Bosons in one-dimensional repulsive potential at finite temperatures}

About this paper \cite{69a},  published in 1969 jointly with C. P. Yang, Yang made the following revealing remark in his {\em Selected Papers} of 1983: ``The rigor with which we had established the Bethe's hypothesis in [[66e]] now paid off. It allowed us to have a firm grip on the quantum numbers $I_1, I_2 \cdots$ of the present problem. The {\em security} generated by such an understanding allowed us to take the next jump, which led to the solution of the finite-temperature problem. It formed, in a sense, a solid platform from which to take off.''~\cite{exp4}  Recently, the model and its solution in this paper have  been realized and confirmed in cold atom experiments \cite{yangyang,van}.

\subsection*{B.1. Flux quantization in superconductors}

In the summer of 1961,  Yang was visiting  Stanford University, where W. M. Fairbank and B. S. Deaver Jr. did an important experiment showing that magnetic flux trapped in a superconducting ring is quantized in units of $hc/2e$. The possibility of such quantization had been conjectured by London and Onsager, but on faulty reasons. With N. Byers, Yang found the correct reason for such quantization~\cite{61c}. They made use of a gauge transformation in their derivation.

\subsection*{B.2. Off-diagonal long-range order}

In 1962, Yang defined the concept of  ODLRO to provide a unified understanding of superfluidity and superconductivity, as well as the origin of flux quantization in superconductors.~\cite{62j} This is a key concept in modern condensed matter physics.  Yang later regarded this paper as  ``a paper I have always been fond of, although it is clearly unfinished''~\cite{exp5}.

In 2006, A. J. Leggett published a book on various quantum condensation phenomena~\cite{leggett}, the preface of which includes the following note: ``I take from the start the viewpoint first  enunciated explicitly by C. N. Yang, namely that one should simply think, in nontechnical  terms, about the behavior of single particles, or pairs of particles, averaged  over the behavior of all the others, or more technically about the one- or two-particle  density matrix.''

\subsection*{C.1. Parity nonconservation in weak interactions}

In a conference  at Stony Brook in 1999, Yang was hailed as ``The Lord of Symmetry''~\cite{jarlslog}. Indeed, if one wants to find a dominant theme in all of Yang's papers and lectures, it would be the concept of symmetry, which in a lecture at UNESCO in 2002 he designated as one of the ``three thematic melodies of 20th century theoretical physics''~\cite{melody}.

Historically symmetry did not play an important role in physics until the advent of quantum theory, which explained the quantum numbers  $l$ and $m$ of atomic spectroscopy in terms of eigenvalues of rotational operators. By early 1950s, all quantum numbers were known to be associated with symmetry operations. Among these parity was related to the reflection operator. Parity conservation was thus considered intuitively attractive, indeed natural and sacred. It was also experimentally extremely useful, especially for analyzing experiments in nuclear physics.

In such an atmosphere it was not surprising that the 1956 paper \cite{56h} proposing experiments to test parity conservation in weak interactions had met with universal disapproval, even ridicule.  It was also not surprising that after C. S. Wu announced her experimental result in early 1957, Yang and Lee's 1956 paper was awarded a Nobel prize later in the same year, setting a record for the speed of recognition that  remains unbroken up today.

\subsection*{C.2. Discrete symmetries of time reversal, charge conjugation and parity}

The preprint of Lee and Yang questioning whether parity is conserved in weak interactions prompted R. Oehme to write to Yang in August 1956 to raise a question about the relationship among the three symmetries, parity (P), charge conjugation (C) and time reversal (T). This led Yang, Lee and Oehme to publish a paper to discuss such relations among violations of these three symmetries~\cite{57e}.  This paper had  decisive impact on  all theoretical analyses on CP violation   later in  1964.

\subsection*{C.3. Theoretical discussions on high energy neutrinos}

In 1960, M. Schwartz  pointed out how one could experimentally study weak interactions at high energies by using neutrino beams~\cite{schwartz}. Lee and Yang then explored theoretically the importance of such experiments in Ref.~\citen{60d}. This paper  had great influences on subsequent neutrino experiments.

J. Steinberger remarked in his 2005 autobiographical book~\cite{steinberger1}: ``The  physics interests of such experiments were tabulated in an accompanying paper by Lee and Yang that proved prophetic. $\cdots$ These processes became topics of extensive experimentation in the following years, as neutrino beams and detectors became increasingly powerful.''

\subsection*{C.4. Phenomenological framework of CP violation}

In 1964, after the experimental discovery of CP violation by J. H. Christenson, J. W. Cronin, V. L. Fitch and R. Turlay, there appeared many theoretical papers on the subject. About these papers, Cronin wrote later in 1993~\cite{cronin}: ``Among all of these theoretical papers of 1964, only two are still quoted today. One   paper is entitled {\em (sic)}  `Phenomenological Analysis of Violation of CP Invariance in Decay of $K^0$ and $\bar{K}^0$' by T. T. Wu and C. N. Yang.'' Cronin also pointed out  that this paper ``has served as a guide to experiments over the past 29 years''~\cite{cronin}.  That Yang chose in 1964 not to speculate on the origin of CP violation, but to concentrate instead on a detailed analysis of possible future experiments  clearly reflects the influence of E. Fermi \cite{fermi}.

Together with Ref.~\citen{57e}, this paper~\cite{64f} defined the formalism and vocabulary of this subject that have been used up to present. Steinberger recalled in his 2005 autobiographical book that it was the Wu-Yang paper that inspired him to measure the major parameters in neutral kaon decays~\cite{steinberger2}.

\subsection*{D.1. Yang-Mill gauge theory}

In two short papers of 1954~\cite{54b,54c}, Yang and R. Mills generalized H. Weyl's Abelian gauge theory to non-Abelian gauge theory. This generalization, together with the ideas of spontaneous symmetry breaking and asymptotic freedom developed later by many authors, led to the Standard Model, which has dominated researches in fundamental physics in all subsequent years.

The motivation of Yang and Mills for making the generalization was clearly stated in Ref.~\citen{54b}, which was a short abstract of Yang's talk  in session M of the 1954 April meeting of American Physical Society in Washington DC. It was submitted, probably before April 1, for publication in the program for that meeting. It says, referring to the conservation of electric charge,  `An important concept in this case is gauge invariance which is closely connected with (1) the equation of motion of the electromagnetic field, (2) the existence of a current density, and (3) the possible interactions between a charged field and the electromagnetic field. We have tried to generalize this concept of gauge invariance to apply to isotopic spin conservation.''

Thus what Yang and Mills were doing was to generalize the close connection between electromagnetic interaction and Abelian gauge invariance to that between a new type of interaction and non-Abelian gauge  invariance. In other words, they were knocking at the door of a principle later called ``Symmetry Dictates Interaction.''~\cite{80b}

\subsection*{D.2. Integral formalism of gauge theory}

When they presented Yang-Mills theory back in 1954, Yang and Mills did not notice the geometric meaning of the gauge field, although they felt it to be a very beautiful theory.

Around 1970, Yang  devoted himself to an integral formalism of gauge field theory and discovered the importance of the nonintegral phase factor, and he realized that gauge field has a deep geometric meaning, which was finally written up in 1974~\cite{74c}.  A few years later, commenting on this paper, Yang wrote~\cite{onmath}:

\begin{quote}
``Most of my physicist colleagues take a utilitarian view about mathematics. Perhaps because of my father's influence, I appreciate mathematics more. I appreciate the value judgment of the mathematicians, and I admire the beauty and power of mathematics: there are ingenuity and intricacy in tactical maneuvers, and breathtaking sweeps in strategic campaigns. And, of course, miracle of miracles, some concepts in mathematics turn out to provide the fundamental structures that govern the physical universe!''
\end{quote}

\subsection*{D.3. Gauge theory and fibre bundles}

In the early 1970s, with the appreciation of the geometric meaning of gauge fields and that the integral formalism of gauge theory is in fact a deep geometrical development, Yang learned from his Stony Brook colleague Jim Simons the rudiments of mathematician's fibre bundle theory. With T. T. Wu,  Yang  finally realized that what physicists call a gauge is what the mathematicians call a principal coordinate bundle~\cite{75c}, and what physicists call potentials are what mathematicians call connections  on principal fiber bundles. They were then able to construct  a ``dictionary'' in this paper, identifying concepts in gauge theory, Abelian or non-Abelian, with those in fibre bundle theory. This dictionary was one of the reasons that mathematicians and theoretical physicists began close collaborations in the last forty years.

M. Atiyah wrote~\cite{atiyah}:  ``From 1977 onwards my interests moved in the direction of gauge theories and the interaction between geometry and physics. $\cdots$ the stimulus in 1977 came from two other sources. On the one hand, Singer told me about the Yang-Mills equations, which through the influences of Yang were just beginning to percolate into mathematical circles.''

\section{Discussions}

As a great master of theoretical physics in the latter half of the 20th century, Chen Ning Yang has a distinctive style and taste. In statistical mechanics and condensed matter physics, as well as in field theory and particle physics, his research works were all characterized by great relevance to experimental facts on the one hand, and great attention to beauty of theoretical form on the other. These characteristics were maintained throughout his long career, from his student days to today: He never followed new fashions, never jumped on bandwagons, but always kept close to his own intuition and to new experimental discoveries. Some of his research works were quickly confirmed by experiments, as in the case of nonconservation of parity in weak interactions. Others had to wait patiently for many years of further developments before their importance were recognized, as in the case of Yang-Mills theory.  That is probably why Yang deeply appreciated the two lines of Tu Fu quoted above.

	It is noteworthy that more than two thirds of the 13 items on the black cube are about the relationship between physical phenomena and algebraic or geometrical symmetries, indicating the central role symmetry plays in Yang's thinking. In fact, in a 1980 paper in {\em Physics Today},  Yang coined the phrase ``Symmetry dictates Interaction''~\cite{80b}. Today it is amply clear: (1) that phrase does concisely   capture the spirit of the main conceptual advance of theoretical physics in the last half century, and (2) that phrase will continue to provide general guidance for future developments in theoretical physics.

	In lectures to students in recent years~\cite{experience,shi}, Yang traced his interests in mathematics and in symmetries to his father's early influence and to Prof. T. Y. Wu's guidance when he was writing his BSc thesis. He emphasized that early interests may form ``seedlings'' which, with sufficient sunlight, nourishment and care, may eventually flower and bear fruit.

\section*{Acknowledgment}
The author is very grateful to Professor Chen Ning Yang for fruitful discussions and for kindly providing the pictures of the black cube.


\begin{thebibliography}{99}
\bibitem{dyson1} F. Dyson, A Conservative Revolutionary, {\em  Int. J. Mod. Phys. A}    {\bf 14}, 1455 (1999).
\bibitem{dyson2} F. Dyson,  Birds and Frogs,  {\it Notices of AMS, 2009,  {\bf 56},  212 (2009)}.
\bibitem{sp} C. N. Yang, {\em Selected Papers 1945-1980 with Commentary}  (W. H. Freeman and Company, 1983) [2005 Edition (World Scientific, Singapore, 2005)].
\bibitem{sp2}  C. N. Yang, {\em Selected Papers II with Commentary}   (World Scientific, Singapore, 2013).
\bibitem{zhang} C. N. Yang, {\em Collected Papers of Chen Ning Yang} (in Chinese), ed. D. Z. Zhang  (East China Normal University, Shanghai, 1998).
\bibitem{shuguangji} C. N. Yang, {\em  C. N. Yang's Dawning Volume} (in Chinese),  ed. F. Weng (World Scientific, Singapore, 2008).
\bibitem{list} Full publication list of C. N. Yang,  \\ http://www.phy.cuhk.edu.hk/people/Yang\_pub\_list.pdf.
\bibitem{goldhaber} A. Goldhaber, R. Shrock, J. Smith, G. Sterman, P. van Niuwenhuizen and W. Weisberger (eds.),  {\em Symmetry and Modern Physics} (World Scientific, Singapore,  2003).
\bibitem{yang70} C. S. Liu and S. T. Yau  (eds.), {\em Chen Ning Yang: A Great Physicist of the Twentieth Century} (International,  Boston, 1997).
\bibitem{yang85} M. L. Ge, C. H. Oh and K. K. Phua (eds.) {\em Proceedings of The Conference in Honore of C. N. Yang's 85th Birthday} (World Scientific, Singapore,  2008).
\bibitem{landau}  I. K. Kikoin,   Landau's ten commandments, in {\em Landau, the Physicist and the Man: Recollections of L.D. Landau}, eds. I. M. Khalatnikov, translated by J. B. Sykes  (Pergamon, Oxford, 1989).
\bibitem{chern} C. N. Yang, {\em On the Chern Class} (in Chinese), 七十年代 ({\em Seventies}), February issue, 1983, Hong Kong;  reprinted in Refs.~\citen{zhang} and  \citen{shuguangji}.
\bibitem{52a} C. N. Yang, The Spontaneous Magnetization of a Two-Dimensional Ising Model, {\em Phys. Rev.}  {\bf 85}, 808 (1952).
\bibitem{52b} C. N. Yang and T. D. Lee, Statistical Theory of Equations of State and Phase Transitions. I. Theory of Condensation, {\em  Phys. Rev.}  {\bf 87}, 404 (1952).
\bibitem{52c} T. D. Lee and C. N. Yang, {\em Statistical Theory of Equations of State and Phase Transitions. II. Lattice Gas and Ising Model}, Phys. Rev.  {\bf 87}, 410 (1952).
\bibitem{57h} T. D. Lee and C. N. Yang, Many-Body Problem in Quantum Mechanics and Quantum Statistical Mechanics, {\em Phys. Rev.} {\bf  105}, 1119 (1957).
\bibitem{57i} T. D. Lee, K. Huang, and C. N. Yang, Eigenvalues and Eigenfunctions of a Bose System of Hard Spheres and Its Low-Temperature Properties, {\em Phys. Rev.} {\bf 106}, 1135 (1957).
\bibitem{57q} K. Huang, T. D. Lee, and C. N. Yang, Quantum Mechanical Many-Body Problem and the Low Temperature Properties of a Bose System of Hard Spheres, Lecture given at the {\em Stevens Conference on the Many-Body Problem}, January 1957,   in {\em The Many-Body Problem}, ed. J. K. Percus ( Wiley-Interscience, New York, 1963), p. 165.
\bibitem{67e} C. N. Yang, Some Exact Rcsults for the Many-Body Problem in One Dimension with Repulsive Delta-Function Interaction, {\em Phys. Rev. Lett.} {\bf 19}, 1312 (1967).
\bibitem{69a}  C. N. Yang and C. P. Yang, Thermodynamics of a One-Dimensional System of Bosons with Repulsive Delta-Function Interaction, {\em  J.  Math. Phys.}  {\bf 10}, 1115 (1969).
\bibitem{61c} N. Byers and C. N. Yang, Theoretical Considerations Concerning Quantized Magnetic Flux in Superconducting Cylinders, {\em  Phys. Rev. Lett.} {\bf 7}, 46 (1961).
\bibitem{62j} C. N. Yang, Concept of Off-Diagonal Long-Range Order and the Quantum Phases of Liquid He and of Superconductors, {\em Rev. Mod. Phys.} {\bf 34}, 694 (1962).
\bibitem{56h} T. D. Lee and C. N. Yang, Question of Parity Conservation in Weak Interactions, {\em  Phys. Rev.}  {\bf  104}, 254 (1956).
\bibitem{57e} T. D. Lee, R. Oehme, and C. N. Yang, Remarks on Possible Noninvariance Under Time Reversal and Charge Conjugation, {\em Phys. Rev.}  {\bf 106}, 340 (1957).
\bibitem{60d} T. D. Lee and C. N. Yang, Theoretical Discussions on Possible High-Energy Neutrino Experiments, {\em  Phys. Rev. Lett.}  {\bf 4}, 307 (1960).
\bibitem{64f} T. T. Wu and C. N. Yang, Phenomenological Analysis of Violation of CP Invariance in Decay of $K^0$ and $\bar{K}^0$, {\em Phys. Rev. Lett.} {\bf 13}, 380 (1964).
\bibitem{54b} C. N. Yang and R. Mills, Isotopic Spin Conservation and a Generalized Gauge Invariance, {\em  Phys. Rev.} {\bf 95}, 631 (1954).
\bibitem{54c} C. N. Yang and R. L. Mills,  Conservation of Isotopic Spin and Isotopic Gauge Invariance, {\em Phys. Rev.} {\bf 96}, 191 (1954).
\bibitem{74c} C. N. Yang, Integral Formalism for Gauge Fields, {\em Phys. Rev. Lett.}  {\bf 33}, 445 (1974).
\bibitem{75c} T. T. Wu and C. N. Yang,  Concept of Nonintegrable Phase Factors and Global Formulation of Gauge Fields, {\em Phys. Rev. D }  {\bf 12}, 3845 (1975).
\bibitem{exp1} F. Dyson, see Ref.~\citen{yang70}, p.~131.
\bibitem{exp2} C. N. Yang, see Ref.~\citen{sp}, p.~12.
\bibitem{exp3} C. N. Yang, see Ref.~\citen{sp}, p.~15.
\bibitem{ruelle} D. Ruelle, {\em Mathematician's Brain} (Princeton University, Princeton, 2007), pp. 91-94.
\bibitem{mayor} J. E. Mayor, The Statistical Mechanics of Condensing Systems. I,   {\em J. Chem. Phys.}  {\bf 5}, 67 (1937).
\bibitem{bf} M. Born and K. Fuchs,  The Statistical Mechanics of Condensing Systems,  {\em Proc. Roy. Soc. (London) A } {\bf 166}, 391 (1938).
\bibitem{altmeyer} A. Altmeyer {\it et al.},  Precision Measurements of Collective Oscillations in the BEC-BCS Crossover,  {\em  Phys. Rev. Lett.}  {\bf 98}, 040401 (2008).
\bibitem{navon} N. Navon {\it et al.},  Dynamics and Thermodynamics of the Low-Temperature Strongly Interacting Bose Gas, {\em Phys. Rev. Lett.}  {\bf 107}, 135301 (2011).
\bibitem{li} B. A.  Li, in Ref.~\citen{yang70}, p.~193.
\bibitem{moritz}  H. Moritz  {\it et al.}, Confinement Induced Molecules in a 1D Fermi Gas, {\em Phys. Rev. Lett.}  {\bf 94}, 210401 (2005).
\bibitem{exp4} C. N. Yang, see Ref.~\citen{sp}, p.~68.
\bibitem{yangyang} T. Kinoshita, T.  Wenger and D. S. Weiss,  Local pair correlations in onedimensional Bose gases,  {\em   Phys. Rev. Lett.}  {\bf 95}, 190406 (2005).
\bibitem{van} A. H. van Amerongen {\it et al.}, Yang-Yang thermodynamics on an atom chip, Phys. Rev. Lett. {\bf 100}, 090402 (2008).
\bibitem{exp5} C. N. Yang, see Ref.~\citen{sp}, p.~54.
\bibitem{leggett} A. J. Leggett, {\em Quantum Liquids}     (Oxford University, Oxford, 2006).
\bibitem{jarlslog}  C. Jarlskog, Speech at {\em  Chen Ning Yang Retirement Symposium}, May 21-22,  1999,  unpublished.
\bibitem{melody} C. N. Yang,  Thematic Melodies of Twentieth Century Theoretical Physics: Quantization， Symmetry and Phase Factor, {\em  Int. J. Mod. Phys. A},    {\bf 18},  3263 (2003);  reprinted in Ref.~\citen{sp2}.
\bibitem{schwartz} M. Schwartz,    Feasibility of Using High-Energy Neutrinos to Study the Weak Interactions, {\em  Phys. Rev. Lett. } {\bf 4}, 306 (1960).
\bibitem{steinberger1} J. Steinberger,  {\em Learning about Particles--50 Privileged Years}  (Springer, Berlin, 2005), p. 87.
\bibitem{cronin} J. W. Cronin, in Ref.~\citen{yang70}, p.~88.
\bibitem{fermi} A. Pais, {\em  Inward Bound}  (Oxford University, Oxford, 1986), p. 533.
\bibitem{steinberger2} J. Steinberger, see Ref.~\citen{steinberger1}, pp. 96-105.
\bibitem{80b} C. N. Yang,  Einstein's Impact on Theoretical Physics,   {\em    Physics Today} {\bf 33} (6), 42 (1980).
\bibitem{onmath} C. N. Yang,  see Ref.~\citen{sp}, p.~74.
\bibitem{atiyah} M. Atiyah,  {\em Collected Works: Volume 5: Gauge Theories} (Oxford University, Oxford, 1988),  p.~1.
\bibitem{experience} C. N.  Yang,  My Experiences of Study and Research  (in Chinese),  {\em Physics}, {\bf 41} (1), 1 (2012).
\bibitem{shi} Y. Shi and Y. Dai,   Discussions with Prof. Chen Ning Yang  (in Chinese),  {\em Physics}, {\bf 40} (8), 491 (2011).
\end{thebibliography}
\end{document}